\documentclass[letterpaper, 10 pt, conference]{ieeeconf}

\IEEEoverridecommandlockouts
\usepackage[T1]{fontenc}
\usepackage{graphicx}
\usepackage{graphics}
\usepackage[caption=false,font=footnotesize]{subfig}
\usepackage{booktabs}
\usepackage{multirow}
\usepackage{makecell}
\usepackage{siunitx}
\usepackage{lmodern}
\usepackage{amsmath}
\usepackage{amssymb}
\usepackage{hyperref}
\usepackage{xurl}
\usepackage{xcolor}
\usepackage{dblfloatfix}

\renewcommand{\texttt}[1]{{\small\bfseries\ttfamily\fontseries{m}\fontshape{n}#1}}

\title{\LARGE \bf
Impact of RTK Augmentation and INS Integration on GNSS Positioning Accuracy and Continuity: A Benchmarking Study on Inland Waterways$^{\star}$}

\author{Yan-Yun Zhang$^{*,1,2}$, Jef Billet$^{1}$, Jan Swevers$^{1,2}$, and Peter Slaets$^{1}$
\thanks{$^{\star}$This is a paper accepted for presentation at the 10th IEEE Conference on Control Technology and Applications (CCTA 2026).~\textcopyright~2026 IEEE. Personal use of this material is permitted. Permission from IEEE must be obtained for all other uses, in any current or future media, including reprinting/republishing this material for advertising or promotional purposes, creating new collective works, for resale or redistribution to servers or lists, or reuse of any copyrighted component of this work in other works.}
\thanks{$^{*}$Corresponding author: Yan-Yun Zhang (e-mail: \href{mailto:yanyun.zhang@kuleuven.be}{yanyun.zhang@kuleuven.be}).}
\thanks{$^{1}$All authors are with the Department of Mechanical Engineering, KU Leuven, 3001 Leuven, Belgium.}%
\thanks{$^{2}$Yan-Yun Zhang and Jan Swevers are also with Flanders Make@KU Leuven, 3001 Leuven, Belgium.}%
}
\begin{document}

\maketitle
\thispagestyle{empty}
\pagestyle{empty}

\begin{abstract}
Real-Time Kinematic (RTK) augmentation and Inertial Navigation System (INS) integration are widely used to improve Global Navigation Satellite System (GNSS) positioning performance. However, on inland waterways, bridges and surrounding structures can degrade satellite visibility and correction availability, causing RTK augmentation loss, and GNSS/INS fusion transients. Since these effects depend on the local environment and sensor configuration, nominal receiver specifications are insufficient, and deployment-specific characterization is required.

This paper presents a benchmarking study of an AsteRx-i3 D Pro+ GNSS/INS receiver installed within the mobile \emph{Sensor Box} developed at KU Leuven. The study combines a real-world bridge-passage case study, static benchmarking, and closed-loop path-following experiments. The static benchmarking evaluates four receiver configurations: standalone GNSS, standalone GNSS with INS integration, RTK-augmented GNSS, and RTK-augmented GNSS with INS integration. The closed-loop experiments use INS-integrated GNSS as the navigation input and compare path-following operational performance with and without RTK augmentation.

Results show that correction loss during bridge passage causes reduced positioning accuracy, increased positioning uncertainty and recovery-induced state jumps exceeding 1\,m. Static benchmarking and closed-loop experiments confirm that RTK augmentation substantially improves positioning precision and uncertainty consistency, while INS integration supports short-term continuity during RTK unavailability but may introduce drift, bias, or transient uncertainty variations. By characterizing the deployment-specific receiver behavior with RTK augmentation and INS integration, this study motivates higher-level state estimation as a necessary next step toward spatially continuous and uncertainty-consistent positioning on inland waterway.
The experimental data are released at: \url{https://doi.org/10.5281/zenodo.20541733} to support reproducibility and further research.
\end{abstract}

\section{INTRODUCTION}\label{sec:introduction}
Inland Waterway Transportation (IWT) is a sustainable and efficient alternative to road and rail transportation, offering a low-emission and high-capacity transport mode. To further increase the competitiveness of IWT, digitalization and automation are being actively pursued, with accurate and robust positioning serving as a fundamental enabler. Accurate positioning is particularly demanding on spatially constrained and infrastructure-dense inland waterways, where even moderate positioning errors can substantially reduce safety margins.

Global Navigation Satellite Systems (GNSS) have become the primary means to meet these positioning demands~\cite{EUSPA_2023_Report}, either as the sole measurement source or as the backbone of multi-sensor localization frameworks~\cite{Hoesch_2023_High}. Modern receivers support multiple Position, Velocity, and Time (PVT) solution modes depending on available correction data, such as:

\begin{itemize}
    \item \texttt{Standalone mode}: The receiver relies on satellite signals alone and typically provides meter-level accuracy under nominal open-sky conditions;
    \item \texttt{Differential mode}: This mode uses code-based corrections from reference stations or networked services and can improve accuracy to the sub-meter level;
    \item \texttt{RTK mode}: Real-Time Kinematic uses carrier-phase-based corrections from a nearby reference station or networked service. Successful integer ambiguity resolution yields \texttt{RTK-fixed mode} with centimeter-level accuracy, while unresolved ambiguities result in \texttt{RTK-float mode} with reduced precision.
\end{itemize}

Among these modes, \texttt{RTK mode} is particularly relevant for accuracy-critical inland navigation as it can achieve centimeter-level positioning accuracy. RTK corrections are encoded according to standardized correction message formats specified by the Radio Technical Commission for Maritime Services (RTCM)~\cite{RTCM_2024_DGNSS} and are commonly transmitted to users over cellular links using the Networked Transport of RTCM via Internet Protocol (NTRIP)~\cite{RTCM_2024_NTRIP}. In Europe, a diverse ecosystem provides such services: (i) commercial providers (e.g., HxGN SmartNet~\cite{HxGN_SmartNet}); (ii) national geodetic agencies (e.g., Belgium's AGN~\cite{AGN}); and (iii) community networks (e.g., RTK2GO~\cite{RTK2GO}). Together with dense reference station coverage and the move toward open public-sector data~\cite{EU_2019_Directive}, this ecosystem has transformed RTK augmentation from a specialized surveying tool into a broadly accessible positioning infrastructure.

However, the operational performance of RTK-augmented GNSS on inland waterways often deviates from ideal conditions. Bridges, lock walls, quay structures, and riverside vegetation can obstruct or reflect satellite signals and interrupt cellular connectivity for NTRIP streams, leading to multipath errors, carrier-phase loss, correction outages, and GNSS PVT mode transitions. A common mitigation is to integrate GNSS with an Inertial Navigation System (INS), which propagates the navigation state using accelerometer and gyroscope measurements. For commercial receivers like the \emph{AsteRx-i3 D Pro+}~\cite{AsteRx-i3D-Pro+} used in this study, this fusion is implemented in firmware, providing a loosely coupled INS-integrated GNSS solution as the primary navigation output. INS integration improves short-term continuity during brief GNSS degradations or outages, but its performance remains dependent on GNSS PVT solution quality and fusion-algorithm integrity~\cite{Groves_2013_Principles}. During GNSS PVT mode transitions, the integrated output may still exhibit spatial discontinuities, drift, or changes in the receiver-reported measurement uncertainties.

Several studies have evaluated GNSS positioning performance with RTK augmentation and INS integration. In the inland waterway context, Yayla et al. benchmarked GNSS positioning with and without INS integration using GPS, Galileo, European Geostationary Navigation Overlay Service (EGNOS), and RTK over repeated runs along a common route~\cite{Yayla_2020_Accuracy}. Their results show that the INS-integrated output largely followed the error characteristics of the underlying GNSS PVT mode. However, the study did not specifically examine the effects of INS integration during mode degradation and recovery, where inertial propagation is expected to be most relevant. Specht evaluated a GNSS/INS system set to operate in \texttt{RTK mode} during inland hydrographic surveys and reported fallback to \texttt{Differential mode} for 30--40\% of the experiment due to limited mobile-network coverage~\cite{Specht_2024_Testing}. The analysis mainly focused on aggregate accuracy and survey-compliance metrics. Beyond waterborne applications, Swaminathan et al. compared differential, RTK, and precise-point-positioning augmentation methods in urban environments, showing strong environment-dependent performance across city-center, underpass, and tunnel scenarios~\cite{Swaminathan_2022_Performance}. Similarly, the analysis did not focus on mode-transition intervals.

These studies mainly address route-level accuracy, survey compliance, or augmentation-method comparison. Less attention has been given to state discontinuities, receiver-reported uncertainties inconsistency, transient behaviors during GNSS degradation and recovery. This missing characterization is important for autonomous navigation systems, where higher-level state estimators are commonly introduced above sensor outputs to improve spatial continuity and statistical consistency~\cite{Bar_2001_Estimation}. The design and tuning of such estimators require a realistic understanding of the deployed receiver, which depends on antenna placement, calibration, firmware implementation, operating environment, and receiver status. Manufacturer specifications and nominal mode-level accuracy claims therefore do not directly translate to a particular system. Deployment-specific benchmarking is consequently needed not only to quantify positioning accuracy, but also to characterize mode-dependent uncertainty behavior and transient state discontinuities relevant to downstream state estimation and control.

This work presents a benchmarking study of the GNSS/INS receiver installed within a mobile \emph{Sensor Box}~\cite{Zhang_2026_Vessel} developed at KU Leuven. The remainder of this paper is organized as follows. Section~\ref{sec:platform} describes the GNSS/INS receiver configuration within the \emph{Sensor Box}. Section~\ref{sec:bridge} analyzes INS-integrated GNSS positioning behavior under signal obscuration during vessel passages under bridges, illustrating real-world mode-transition challenges. Section~\ref{sec:static} presents the static benchmarking results for four receiver configurations: GNSS-only and GNSS/INS-integrated solutions under both \texttt{Standalone} and \texttt{RTK-fixed} modes. These two modes are used as representative lower- and upper-performance cases. Section~\ref{sec:dynamic} evaluates closed-loop path-following performance using the INS-integrated GNSS solution as the primary navigation input. Finally, Section~\ref{sec:conclusions} summarizes the findings and discusses their implications for autonomous navigation on inland waterways.

\begin{figure}[t!]
    \centering
    \includegraphics[width=1.0\linewidth]{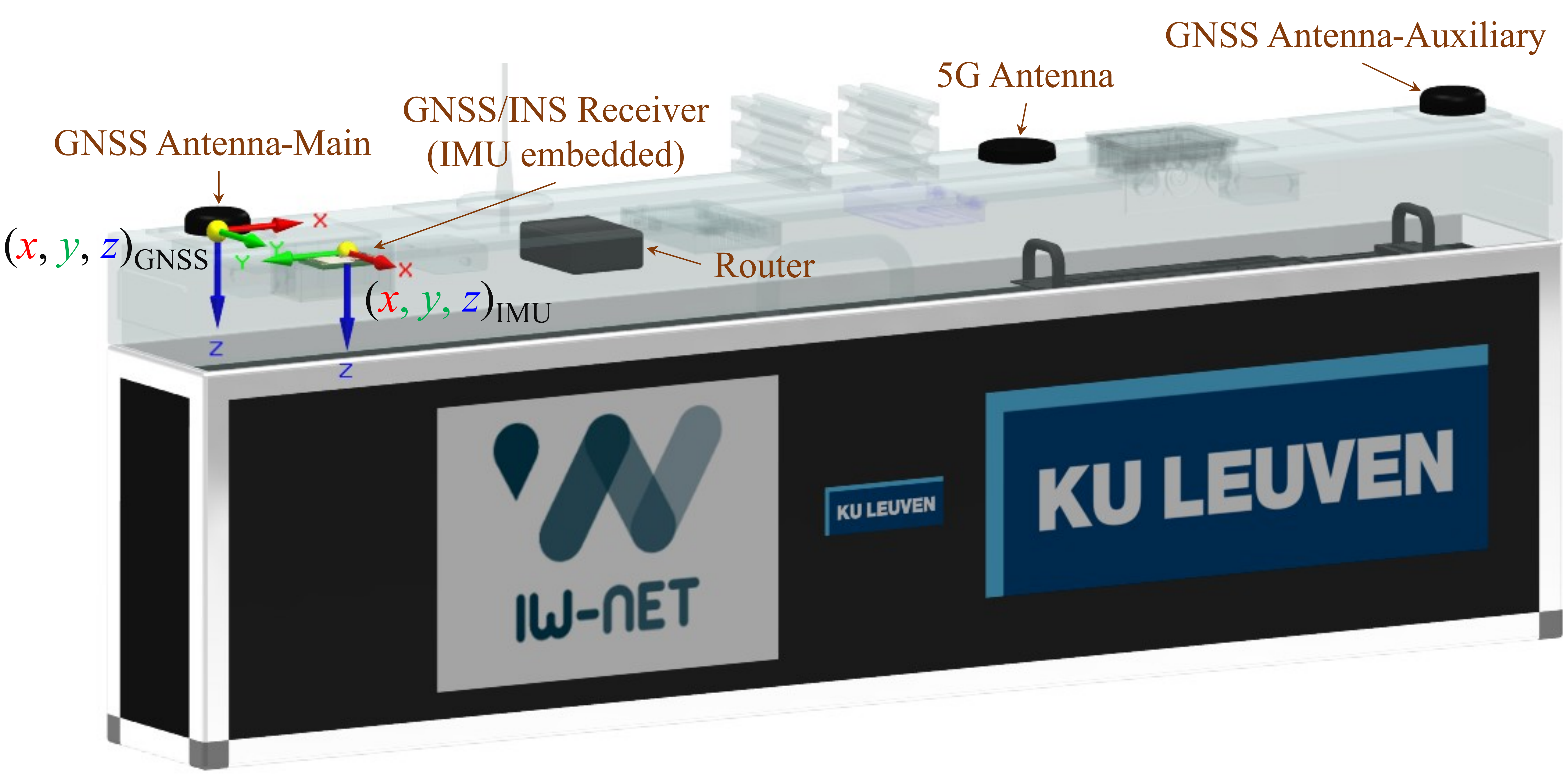}
    \caption{The \emph{Sensor Box} with its key GNSS/INS components annotated and sensor frames specified.}\label{fig:sensor_box}
\end{figure}

\begin{figure*}[b!]
    \centering
    \includegraphics[width=1.0\textwidth]{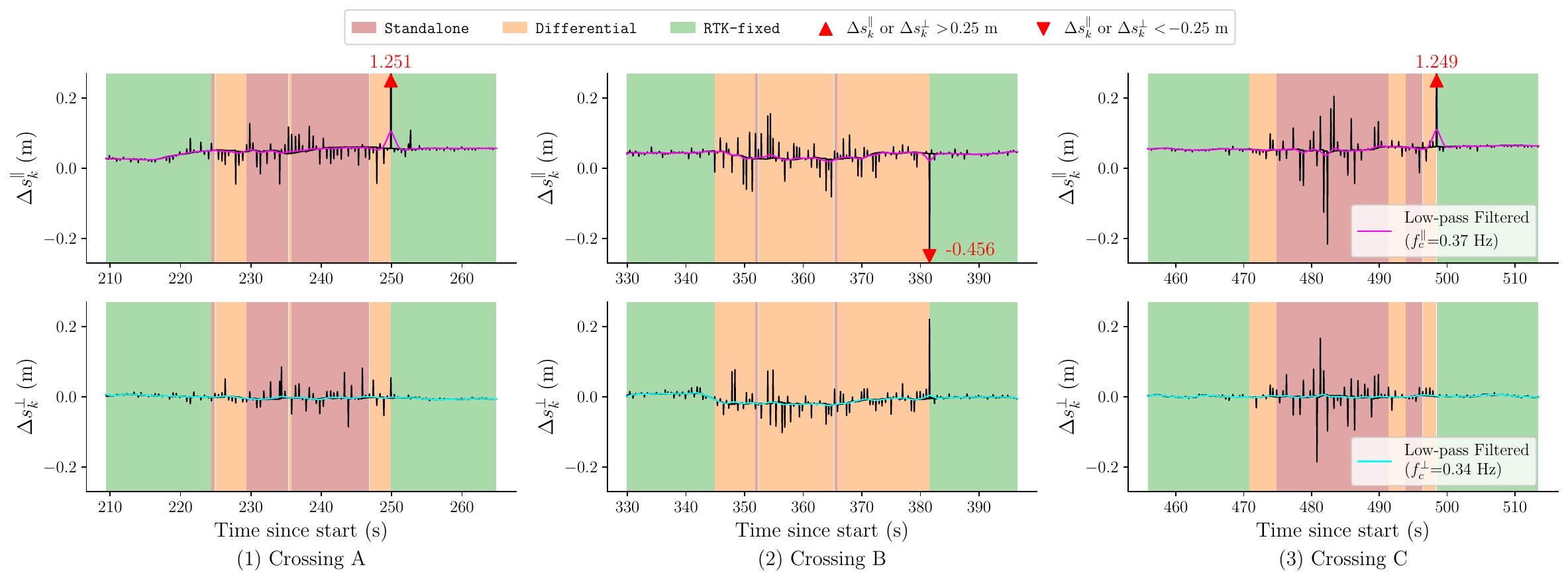}
    \addtocounter{figure}{1}
    \caption{Analysis of along-track and cross-track position differences ($\Delta s_k^{\parallel}$, $\Delta s_k^{\perp}$) during bridge-crossing intervals A, B, and C, with shaded regions indicating the GNSS PVT solution mode.}\label{fig:gnss_trakerror}
    \addtocounter{figure}{-1}
\end{figure*}

\begin{figure}[tb!]
    \centering
    \includegraphics[width=1.0\linewidth]{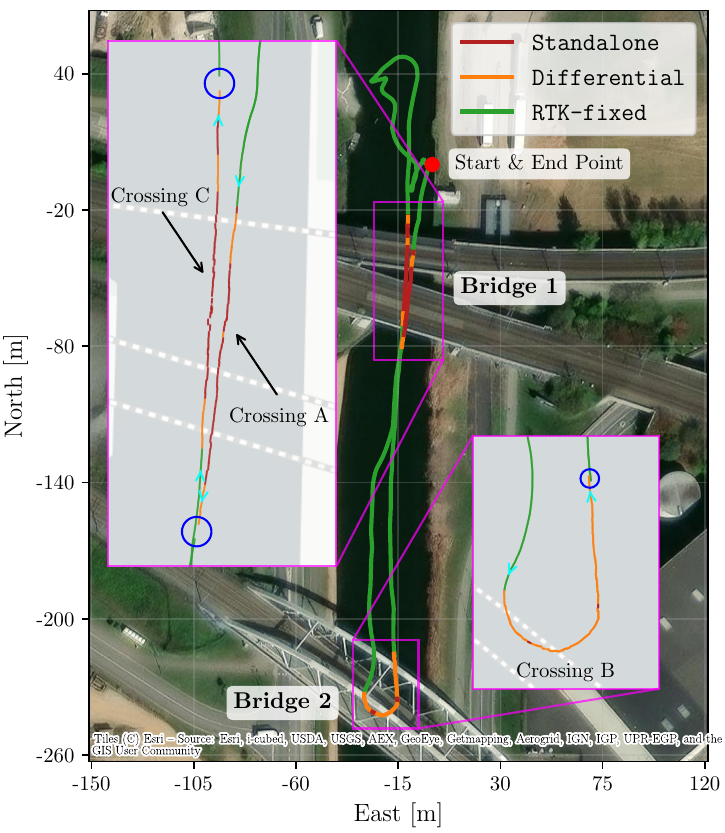}
    \addtocounter{figure}{-1}
    \caption{Recorded path measurements during a canal voyage under bridges, color-coded by the GNSS PVT solution mode.}\label{fig:gnss_trajectory}
    \addtocounter{figure}{1}
\end{figure}

\section{PLATFORM SPECIFICATION}\label{sec:platform}
Fig.~\ref{fig:sensor_box} illustrates the \emph{Sensor Box} used in this study, with key GNSS/INS components annotated. The core component is the \emph{AsteRx-i3 D Pro+}, a compact GNSS/INS receiver board with an integrated Inertial Measurement Unit (IMU). It supports quad-constellation, multi-frequency, all-in-view satellite tracking and accepts external RTCM correction messages via NTRIP for high-precision positioning. Two GNSS antennas are installed with a 1.54\,m baseline to support heading and attitude estimation. 4G/5G connectivity for correction data reception is provided by a dual-SIM router and a dedicated 5G antenna to enhance communication robustness. The receiver provides two primary navigation data outputs:

\begin{itemize}
    \item \texttt{raw} data stream: The GNSS PVT solution.
    \item \texttt{fused} data stream: The INS-integrated solution, computed internally by the receiver using a proprietary Extended Kalman Filter (EKF) that fuses the GNSS PVT solution with IMU measurements.
\end{itemize}

\noindent The \texttt{raw} stream reports the receiver's GNSS PVT solution directly at the configured GNSS update rate. In contrast, the \texttt{fused} stream is available at a higher rate because the internal EKF propagates the navigation state between successive GNSS PVT updates using high-rate IMU measurements. In this study, the \texttt{raw} and \texttt{fused} streams were logged at 2\,Hz and 20\,Hz, respectively. A lever-arm calibration was performed to account for the spatial offset between the sensor frames ${(x, y, z)}_{\text{GNSS}}$ and ${(x, y, z)}_{\text{IMU}}$. Both streams were expressed in the same coordinate frame, ${(x, y, z)}_{\text{GNSS}}$, with the origin fixed at the main GNSS antenna phase center.

For the benchmarking study, the following data fields were extracted from both the \texttt{raw} and \texttt{fused} data streams: the GNSS PVT solution mode (e.g., \texttt{Standalone}, \texttt{Differential}, \texttt{RTK-fixed}, \texttt{RTK-float}), the geodetic position (latitude $\phi$, longitude $\lambda$), and the receiver-reported horizontal measurement uncertainties (standard deviations $\sigma_\phi$ and $\sigma_\lambda$). Note that these uncertainties are internally computed proprietary estimates provided by the receiver, rather than externally validated accuracy metrics. For analysis, all geodetic coordinates were transformed into a local North-East-Down (NED) coordinate frame. Positions are expressed by their North and East components (in meters), and the associated standard deviations are correspondingly transformed and denoted as $\sigma_{\mathrm{N}}$ and $\sigma_{\mathrm{E}}$ (in meters).

\section{SIGNAL OBSCURATION CHALLENGE}\label{sec:bridge}
The GNSS/INS receiver can provide centimeter-level positioning accuracy when operating in \texttt{RTK-fixed mode} under favorable conditions. However, its performance can degrade drastically when satellite signals or correction data are obstructed, for example when a vessel passes under bridges. To illustrate this challenge, Fig.~\ref{fig:gnss_trajectory} shows a recorded trajectory during which the \emph{Sensor Box} was deployed on the research vessel \emph{Maverick}~\cite{Zhang_2023_Design}, sailing along a Belgian canal. The trajectory is plotted using the \texttt{fused} data stream and color-coded according to the GNSS PVT mode at each epoch.

The receiver operated predominantly in \texttt{RTK-fixed mode}; however, automatic transitions to \texttt{Differential} and \texttt{Standalone mode} occurred near and under the bridges. The enlarged views highlight three bridge-crossing intervals, denoted as Crossings A--C. The direction of travel and sequence of mode transitions are indicated by cyan arrows along the path. During these intervals, reduced path smoothness and abrupt position discontinuities upon exiting the bridges can be observed. These discontinuities, highlighted by blue circles in Fig.~\ref{fig:gnss_trajectory}, occur close to transitions back to \texttt{RTK-fixed mode}, indicating that restoring RTK augmentation itself can introduce state jumps.

Because the receiver-reported uncertainties were not logged during this initial field trial, the trajectory was further analyzed using sequential position differences between consecutive epochs $k-1$ and $k$. The inter-epoch position increments in East and North directions, denoted as $\Delta x_k$ and $\Delta y_k$, were projected into along-track and cross-track components as:

\begin{equation}
\Delta s_k^{\parallel} = \Delta x_k \sin(\psi_k) + \Delta y_k \cos(\psi_k)
\end{equation}
\begin{equation}
\Delta s_k^{\perp} = \Delta x_k \cos(\psi_k) - \Delta y_k \sin(\psi_k)
\end{equation}

\noindent where $\psi_k$ denotes the vessel heading at epoch $k$, obtained from the \texttt{fused} data.

Fig.~\ref{fig:gnss_trakerror} presents the resulting $\Delta s_k^{\parallel}$ and $\Delta s_k^{\perp}$ during the three bridge-crossing intervals, color-coded according to the GNSS PVT solution mode. These time series contain both low-frequency components, which are consistent with the vessel's maneuvering motion, and higher-frequency components, which are more likely associated with measurement noise and positioning errors. To aid interpretation, zero-phase low-pass filters were applied to both $\Delta s_k^{\parallel}$ and $\Delta s_k^{\perp}$. The cut-off frequencies, $f_c^{\parallel}$ and $f_c^{\perp}$, were selected slightly above the estimated maneuvering frequency range of the vessel, so that motion-consistent variations were retained while faster fluctuations were attenuated.

The results show that transitions from \texttt{RTK-fixed} to lower-accuracy modes are accompanied by increased high-frequency fluctuations. When the receiver returns to \texttt{RTK-fixed}, abrupt inter-epoch position differences are observed, especially in the along-track direction. In Crossings A and C, the along-track jumps exceed 1\,m, while Crossing B shows a smaller jump in the opposite direction. These events correspond to the path discontinuities observed in Fig.~\ref{fig:gnss_trajectory}. These jumps are likely caused by an abrupt correction of accumulated positioning errors when RTK augmentation is restored and the receiver returns to \texttt{RTK-fixed mode}. Therefore, although INS integration helps maintain short-term continuity during temporary RTK unavailability, the INS-integrated GNSS solution may still exhibit spatial discontinuities during mode recovery. For autonomous navigation, this indicates that GNSS PVT mode transitions and recovery-induced state jumps should be monitored by a higher-level state estimation or supervision layer before the measurements are used directly for control.

\section{STATIC BENCHMARKING}\label{sec:static}
Static benchmarking provides a controlled baseline for evaluating how RTK augmentation and INS integration affect the positioning behavior of the receiver. In this setting, the observed position scatter mainly reflects differences among the tested solution modes, excluding the influence of platform motion.

\begin{figure}[tb!]
    \centering
    \includegraphics[width=1.0\linewidth]{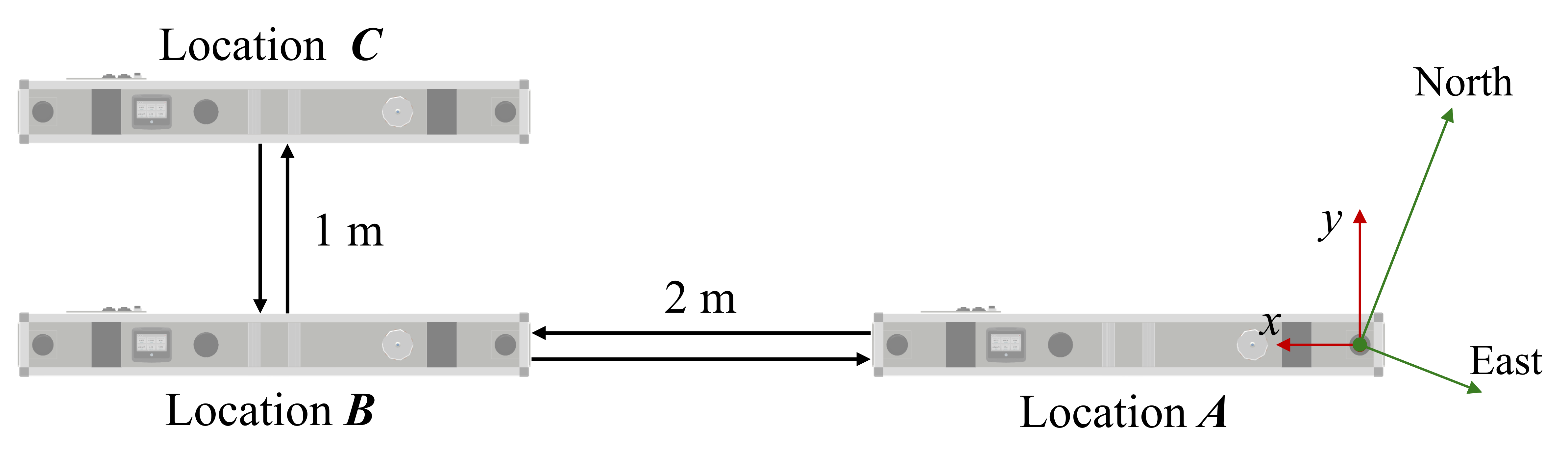}
    \caption{Static benchmarking experimental layout showing three predefined locations arranged in an L-shape, with the local NED coordinate frame annotated.}\label{fig:lshape}
\end{figure}

The \emph{Sensor Box} was placed at three predefined locations, denoted as \textbf{\textit{A}}, \textbf{\textit{B}}, and \textbf{\textit{C}}, arranged in an L-shape as shown in Fig.~\ref{fig:lshape}. At each location, the box remained stationary for approximately 200\,s before being moved to the next one. The full sequence was \textbf{\textit{A}} $\rightarrow$ \textbf{\textit{B}} $\rightarrow$ \textbf{\textit{C}} $\rightarrow$ \textbf{\textit{B}} $\rightarrow$ \textbf{\textit{A}}. Two runs were conducted: one without RTK augmentation, corresponding to \texttt{Standalone mode}, and one with RTK augmentation, corresponding to \texttt{RTK-fixed mode}. During both runs, the \texttt{raw} and the \texttt{fused} data streams were logged simultaneously.

\begin{figure*}[tb!]
    \centering
    \includegraphics[width=1.0\textwidth]{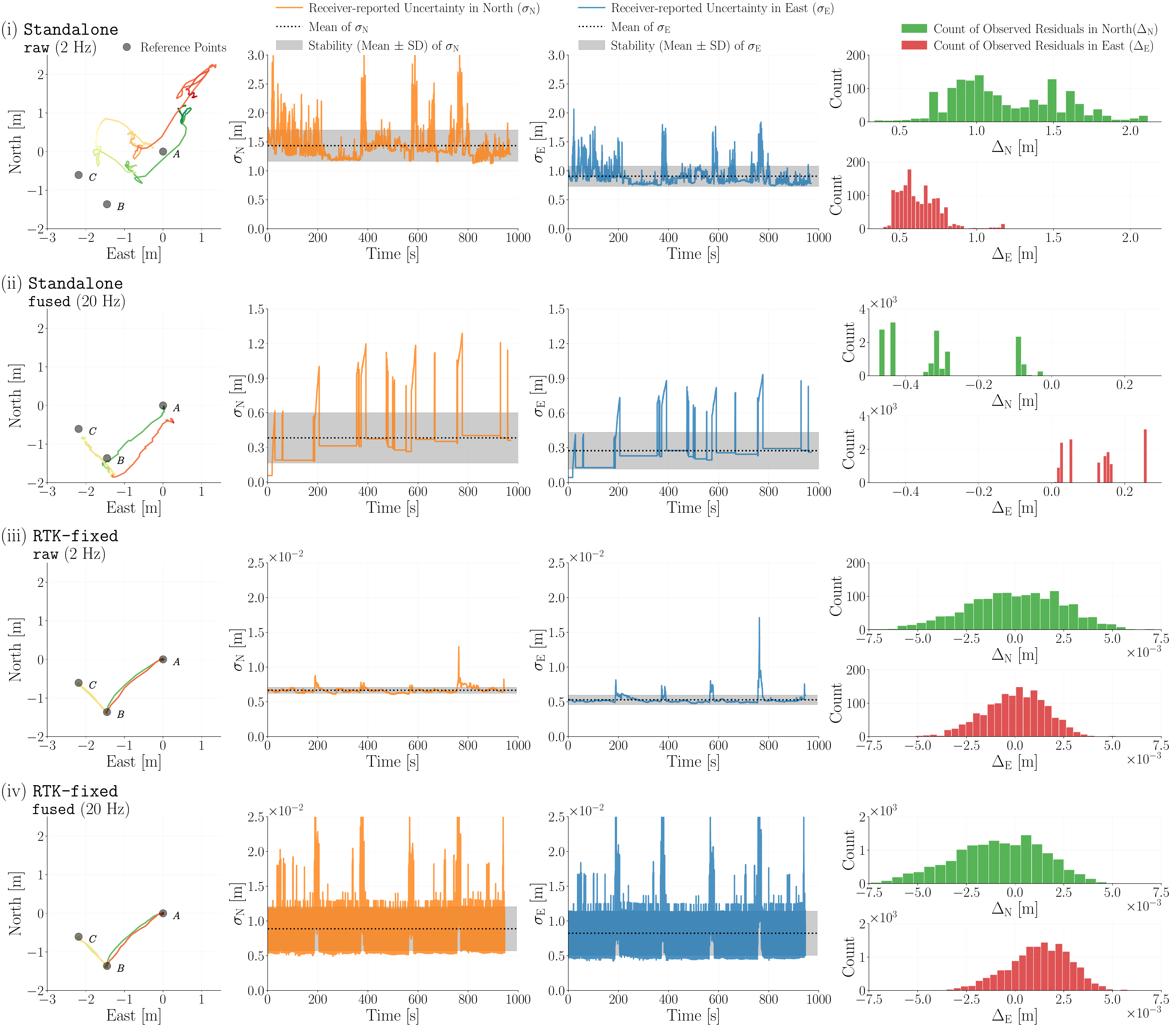}
    \caption{Static benchmarking results across four evaluation cases. From left to right, the columns present the measured paths, receiver-reported uncertainties in North ($\sigma_\mathrm{N}$) and East ($\sigma_\mathrm{E}$) directions, and distributions of the observed residuals ($\Delta_\mathrm{N}$ and $\Delta_\mathrm{E}$).}\label{fig:static_benchmark}
\end{figure*}

\renewcommand{\texttt}[1]{{\footnotesize\bfseries\ttfamily\fontseries{m}\fontshape{n}#1}}

\begin{table*}[tb!]
\vspace{6pt}
\caption{Statistics for receiver-reported uncertainties and residuals relative to the RTK-derived reference coordinates.}\label{tab:static_benchmark_uncertainty}
\centering
\begin{tabular*}{0.98\textwidth}{@{\extracolsep{\fill}} l c c c c c c c @{}}
\toprule
Case & Data & Mean [\unit{m}] & SD [\unit{m}] & RMS [\unit{m}] & P68$^*$ [\unit{m}] & P95$^*$ [\unit{m}] & Outl. [\unit{\percent}] \\
\midrule
\multirow{4}{*}{\makecell[l]{(i) \texttt{Standalone}\\\phantom{(i) }\texttt{raw} (2\,Hz)}} 
    & $\sigma_{\mathrm{N}}$ & 1.44 & 0.27 & 1.46 & 1.45 & 1.98 & 2.17 \\
    & $\sigma_{\mathrm{E}}$ & 0.91 & 0.18 & 0.92 & 0.90 & 1.28 & 2.48 \\
    & $\Delta_{\mathrm{N}}$ & 1.19 & 0.35 & 1.25 & 1.40 & 1.80 & 0.00 \\
    & $\Delta_{\mathrm{E}}$ & 0.63 & 0.13 & 0.64 & 0.67 & 0.84 & 2.06 \\
\addlinespace
\multirow{4}{*}{\makecell[l]{(ii) \texttt{Standalone}\\\phantom{(ii) }\texttt{fused} (20\,Hz)}}
    & $\sigma_{\mathrm{N}}$ & 0.38 & 0.22 & 0.44 & 0.37 & 1.00 & 4.01 \\
    & $\sigma_{\mathrm{E}}$ & 0.27 & 0.16 & 0.32 & 0.28 & 0.72 & 3.78 \\
    & $\Delta_{\mathrm{N}}$ & -0.31 & 0.14 & 0.34 & 0.43 & 0.46 & 0.00 \\
    & $\Delta_{\mathrm{E}}$ & 0.13 & 0.08 & 0.15 & 0.15 & 0.26 & 0.00 \\
\addlinespace
\multirow{4}{*}{\makecell[l]{(iii) \texttt{RTK-fixed}\\\phantom{(iii) }\texttt{raw} (2\,Hz)}}
    & $\sigma_{\mathrm{N}}$ & \num{6.67e-3} & \num{4.34e-4} & \num{6.68e-3} & \num{6.71e-3} & \num{7.47e-3} & 1.22 \\
    & $\sigma_{\mathrm{E}}$ & \num{5.29e-3} & \num{6.91e-4} & \num{5.33e-3} & \num{5.20e-3} & \num{6.18e-3} & 2.27 \\
    & ${\Delta_{\mathrm{N}}}^{**}$ & -- & \num{2.44e-3} & \num{2.44e-3} & \num{2.44e-3} & \num{4.73e-3} & 0.07 \\
    & ${\Delta_{\mathrm{E}}}^{**}$ & -- & \num{1.59e-3} & \num{1.59e-3} & \num{1.60e-3} & \num{3.06e-3} & 0.40 \\
\addlinespace
\multirow{4}{*}{\makecell[l]{(iv) \texttt{RTK-fixed}\\\phantom{(iv) }\texttt{fused} (20\,Hz)}}
    & $\sigma_{\mathrm{N}}$ & \num{8.88e-3} & \num{3.17e-3} & \num{9.43e-3} & \num{9.82e-3} & \num{1.42e-2} & 1.94 \\
    & $\sigma_{\mathrm{E}}$ & \num{8.24e-3} & \num{3.17e-3} & \num{8.83e-3} & \num{9.29e-3} & \num{1.33e-2} & 1.70 \\
    & $\Delta_{\mathrm{N}}$ & \num{-8.87e-4} & \num{2.34e-3} & \num{2.51e-3} & \num{2.45e-3} & \num{5.07e-3} & 0.07 \\
    & $\Delta_{\mathrm{E}}$ & \num{1.23e-3} & \num{1.59e-3} & \num{2.01e-3} & \num{2.19e-3} & \num{3.63e-3} & 0.26 \\
\bottomrule
\end{tabular*}
\par\vspace{2pt}
\begin{minipage}{0.95\textwidth}
\raggedright
\footnotesize{$^*$Percentiles for $\Delta_{\mathrm{N}}$ and $\Delta_{\mathrm{E}}$ are computed on absolute values.}\\
\footnotesize{$^{**}$Statistics for this quantity are self-referenced, as Case (iii) was used to define the RTK-derived reference coordinates.}
\end{minipage}
\end{table*}

\renewcommand{\texttt}[1]{{\small\bfseries\ttfamily\fontseries{m}\fontshape{n}#1}}

No independent geodetic survey of the three locations was available. Therefore, the reference coordinates were defined internally from the experiment. For each location, the reference position was obtained by averaging the measurements from the \texttt{RTK-fixed} \texttt{raw} data stream during the corresponding stationary period. This case was selected because it provided the lowest receiver-reported uncertainties among the tested configurations. The observed residuals in North and East directions, denoted as $\Delta_{\mathrm{N}}$ and $\Delta_{\mathrm{E}}$, were then computed relative to these RTK-derived reference coordinates. They should therefore be interpreted as relative deviations, rather than as absolute errors with respect to an independently surveyed ground truth.

Fig.~\ref{fig:static_benchmark} summarizes the static benchmarking results across the four cases. The first column presents the measured paths, with the RTK-derived reference coordinates as defined above marked as black dots. The second and third columns show the time series of the receiver-reported uncertainties in the North ($\sigma_{\mathrm{N}}$) and East ($\sigma_{\mathrm{E}}$) directions, respectively. The last column displays histograms of the observed residuals in the North ($\Delta_{\mathrm{N}}$) and East ($\Delta_{\mathrm{E}}$) directions.

\paragraph*{Case (i) \texttt{Standalone mode}, \texttt{raw} data stream} This case exhibits the poorest performance, with the measured positions showing clear meter-level bias and large dispersion around the reference locations. At the start of the run, an initial offset from location \textbf{\textit{A}} can be observed, reflecting the absolute bias of the instantaneous GNSS PVT solution under \texttt{Standalone mode}. The receiver-reported uncertainties are large and highly variable, indicating reduced internal confidence and inconsistent uncertainty representation. The residual distributions are right-skewed and non-Gaussian, with errors frequently exceeding 1.5\,m, indicating that the GNSS solution without RTK augmentation and INS integration cannot be modeled as a simple zero-mean measurement source in this deployment.

\paragraph*{Case (ii) \texttt{Standalone mode}, \texttt{fused} data stream}
This case shows significantly improved performance compared with \textit{Case (i)}. The measured path starts much closer to the reference, likely because the internal GNSS/INS filter retained information from previously acquired high-accuracy positioning solutions. Over time, the measured path gradually departs from the reference locations, consistent with increasing reliance on inertial propagation and lower-accuracy GNSS updates after RTK augmentation is no longer available. The residual distributions are more compact than in \textit{Case (i)}, and the receiver-reported uncertainties are also reduced. Nevertheless, the uncertainties remain variable, with intermittent sharp increases, indicating that the uncertainty representation of the INS-integrated solution is not consistently stable under this transition condition. Therefore, this result should be interpreted as a transition-related special case rather than as the general performance of an INS-integrated GNSS solution without RTK augmentation. This behavior is relevant to practical operation, where the receiver may switch from \texttt{RTK-fixed} to \texttt{Standalone mode} near bridges or other obstructions. In such situations, the INS-integrated solution may remain operational, but its receiver-reported uncertainties can exhibit transient peaks and should not be used directly in downstream applications without additional supervision.

\paragraph*{Case (iii) \texttt{RTK-fixed mode}, \texttt{raw} data stream} This case achieves the best overall performance. The receiver-reported uncertainties are low and stable, reflecting high internal confidence with RTK augmentation. Since this case was used to define the internal reference coordinates, its residuals are centered around zero by construction. Nevertheless, the narrow and approximately Gaussian residual distributions confirm that the RTK-augmented GNSS PVT solution is highly stable and repeatable under the tested open-sky conditions.

\paragraph*{Case (iv) \texttt{RTK-fixed mode}, \texttt{fused} data stream} This case provides performance comparable to \textit{Case (iii)}. However, the receiver-reported uncertainties are slightly higher and more variable than those of \textit{Case (iii)}. The residual histograms remain approximately Gaussian, but show a small offset from zero relative to the RTK-derived reference coordinates. This suggests that, when RTK augmentation is continuously available, INS integration does not necessarily further improve the solution. Instead, the INS-integrated solution also reflects IMU noise, sensor-frame alignment errors, and the receiver's proprietary fusion algorithm.

Notably, both $\sigma$ and $\Delta$ are generally smaller in the East direction than in the North direction across all evaluated cases. This suggests anisotropic horizontal positioning behavior, which is expected because GNSS measurement errors are projected through the satellite geometry into the local North-East frame, resulting in direction-dependent horizontal errors.

Furthermore, Table~\ref{tab:static_benchmark_uncertainty} reports the mean, Root Mean Square (RMS), Standard Deviation (SD), 68th and 95th percentiles (P68 and P95), and the outlier percentage (Outl.), defined using a Z-score threshold of 3. For $\sigma_{\mathrm{N}}$ and $\sigma_{\mathrm{E}}$, these statistics characterize the typical reported uncertainty level, its temporal stability, the upper range of reported uncertainty values, and the occurrence of outliers. For $\Delta_{\mathrm{N}}$ and $\Delta_{\mathrm{E}}$, these statistics characterize systematic offset, residual magnitude, empirical dispersion, and the occurrence of outliers.

Overall, the static benchmarking shows that RTK augmentation is the dominant factor in reducing bias and scatter, changing the residual behavior from meter-level, biased, non-Gaussian distributions to centimeter-level, near-Gaussian distributions. INS integration mainly affects solution continuity and uncertainty variability, rather than necessarily improving accuracy under all situations. In \textit{Case (ii)}, the apparent improvement should be interpreted as a transition-related behavior influenced by previously available high-accuracy positioning. Across the four cases, the receiver-reported uncertainties are qualitatively consistent with the observed residual trends, but they should not be interpreted as directly validated accuracy bounds. These results indicate that receiver-reported uncertainties should be interpreted in a configuration- and mode-dependent manner, rather than being used as a single fixed covariance assumption.

\section{CLOSED-LOOP PATH-FOLLOWING EXPERIMENTS}\label{sec:dynamic}

\begingroup
\renewcommand{\texttt}[1]{{\footnotesize\bfseries\ttfamily\fontseries{m}\fontshape{n}#1}}
\begin{figure}[t!]
    \centering
    \subfloat[\texttt{Standalone mode}]{%
        \includegraphics[width=1.0\linewidth]{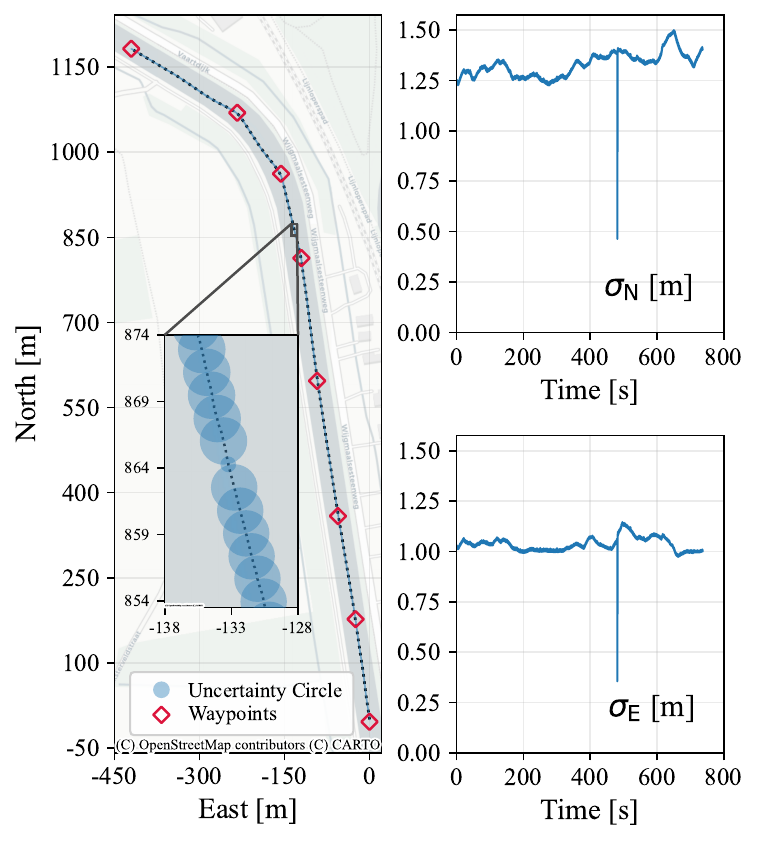}\label{fig:dynamic_test_standalone}
    }\\
    \subfloat[\texttt{RTK-fixed mode}]{%
        \includegraphics[width=1.0\linewidth]{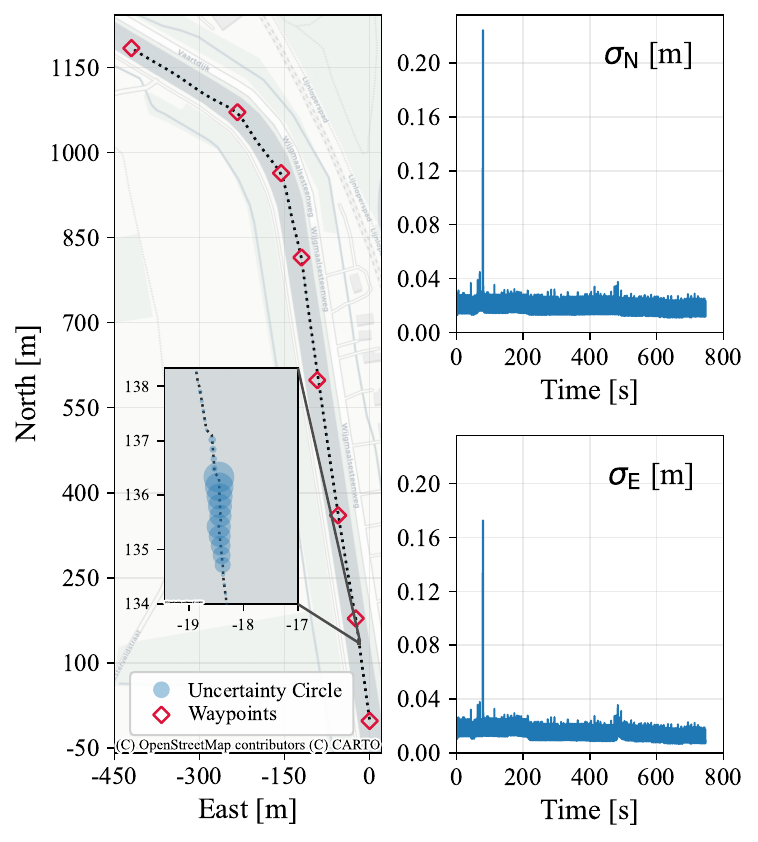}\label{fig:dynamic_test_rtkfixed}
    }
    \caption{Closed-loop path-following experiments using the INS-integrated GNSS solution (\texttt{fused} data stream) as navigation input, comparing \texttt{Standalone} and \texttt{RTK-fixed} positioning modes.}\label{fig:dynamic_test}
\end{figure}
\endgroup

Closed-loop path-following experiments were conducted to evaluate the GNSS/INS receiver under realistic navigation conditions. In these experiments, the \emph{Maverick} followed predefined waypoints at a nominal speed of approximately 1\,m/\,s using the standard Line-of-Sight (LoS) guidance law~\cite{Fossen_2022_Line}. A Proportional-Integral-Derivative (PID) controller regulated the propeller azimuth angles, while the propeller shaft speed was kept constant. The controller used only the \texttt{fused} GNSS/INS data stream as the navigation input. The \texttt{raw} data stream was not used for comparison in this section because its 2\,Hz update rate provides limited temporal resolution for the small-vessel, narrow-canal scenario considered here.

Two runs were conducted: one with the receiver operating in \texttt{Standalone mode} and one in \texttt{RTK-fixed mode}. Fig.~\ref{fig:dynamic_test} summarizes the results. The executed paths are overlaid on the map, with uncertainty circles representing the receiver-reported horizontal uncertainty radius, computed as $\sigma_r=\sqrt{\sigma_{\mathrm{N}}^2+\sigma_{\mathrm{E}}^2}$. For visualization clarity, these circles are plotted at 10\,Hz in Fig.~\ref{fig:dynamic_test_standalone} and at 1\,Hz in Fig.~\ref{fig:dynamic_test_rtkfixed}. The accompanying time series of $\sigma_{\mathrm{N}}$ and $\sigma_{\mathrm{E}}$ provide a clear quantitative view of the receiver-reported uncertainties.

In \texttt{Standalone mode}, the vessel successfully follows the waypoint sequence, but the receiver-reported uncertainties remain at the meter-level, with $\sigma_{\mathrm{N}}$ and $\sigma_{\mathrm{E}}$ generally above 1\,m. This shows that the INS-integrated solution can provide sufficient temporal continuity for waypoint following under the tested conditions. However, the successful path-following result should not be interpreted as evidence of high positioning accuracy. Rather, the application can tolerate the available positioning quality because the tested path has sufficient clearance and does not require centimeter-level precision. A sharp dip appears in the uncertainty time series during receiver internal clock resynchronization. Although the reported uncertainty temporarily decreases, there is no independent evidence that the actual positioning accuracy improves at that moment. This indicates that transient changes in receiver-reported uncertainty may reflect internal numerical behavior rather than a physically verifiable improvement in positioning performance.

In \texttt{RTK-fixed mode}, the uncertainty circles are much smaller and the receiver-reported uncertainties remain mostly at the centimeter-level. A temporary spike appears when the receiver loses integer ambiguity resolution and falls back to \texttt{RTK-float mode}. This spike is consistent with a real degradation in GNSS correction quality and correctly reflects reduced positioning confidence.

Both runs demonstrate successful path following using the INS-integrated GNSS solution. However, RTK augmentation provides a substantially smaller and more stable receiver-reported uncertainty envelope, which is important for navigation in constrained environments. The closed-loop experiments further show that both the position output and the associated uncertainty estimates from the receiver should be interpreted together with its operational context. A higher-level estimator or supervisory layer is therefore needed to monitor GNSS PVT mode transitions, detect inconsistent uncertainty behaviors, and provide spatially continuous navigation states and measurement covariance information to downstream applications.

\addtolength{\textheight}{-0.2cm}

\section{CONCLUSIONS}\label{sec:conclusions}
This paper presented a deployment-oriented characterization of a GNSS/INS receiver, focusing on mode-dependent positioning behavior, receiver-reported uncertainty, and closed-loop navigation relevance.

The bridge-crossing analysis showed that temporary RTK degradation can increase high-frequency positioning fluctuations and lead to recovery-induced state jumps, with along-track jumps exceeding 1\,m in the presented case. INS integration improved short-term continuity but did not fully eliminate discontinuities during RTK recovery.

The static benchmarking further quantified the effects of RTK augmentation and INS integration on positioning scatter and receiver-reported uncertainty. RTK augmentation was found to be the dominant factor in improving positioning quality, changing the residual behavior from biased, meter-level, non-Gaussian distributions in \texttt{Standalone mode} to centimeter-level, near-Gaussian distributions in \texttt{RTK-fixed mode}. INS integration mainly affected temporal continuity and uncertainty behavior rather than uniformly improving accuracy. The results also showed that receiver-reported uncertainties should be interpreted in a configuration- and mode-dependent manner, rather than being treated as fixed covariance values.

The closed-loop path-following experiments confirmed that INS-integrated GNSS can support waypoint tracking under both \texttt{Standalone} and \texttt{RTK-fixed} operation in the tested canal environment. However, successful path following in \texttt{Standalone mode} should not be interpreted as evidence of high positioning accuracy, since the controller could tolerate meter-level uncertainty under the available clearance. RTK augmentation provided a much smaller and more stable uncertainty envelope, which is essential for more constrained maneuvers. Transient dips or spikes in the reported uncertainty further indicate that receiver outputs should be interpreted together with PVT mode, correction availability, and operational context.

Overall, the results show that firmware-level RTK augmentation and INS integration provide substantial benefits, but they do not remove the need for higher-level supervision. Autonomous inland vessels should therefore incorporate supervisory frameworks that monitor GNSS PVT mode transitions, account for mode-dependent measurement quality, detect inconsistent uncertainty behavior, and maintain spatially continuous navigation states for downstream state estimation and control. The empirical characterization presented in this study provides a basis for designing such estimator measurement models and supervisory logic for GNSS/INS-based navigation in infrastructure-dense inland waterways.

\section*{ACKNOWLEDGMENT}
The authors would like to thank Zhongbi Luo for his contribution in data collection. Additionally, the authors acknowledge the use of generative AI tools for assistance in language editing. All scientific interpretations remain the sole responsibility of the authors.

\bibliographystyle{IEEEtran}
\bibliography{ref}

\end{document}